\documentclass[pra,aps,twocolumn]{revtex4}
\usepackage{graphicx}\usepackage{amsmath}

\newcommand{\be}{\begin{equation}}\newcommand{\ee}{\end{equation}}
\newcommand{\bea}{\begin{eqnarray}}\newcommand{\eea}{\end{eqnarray}}
\newcommand{\nn}{\nonumber}

\newcommand{\ep}{\varepsilon}
\renewcommand{\phi}{\varphi}
\newcommand{\Ref}[1]{(\ref{#1})}

\newcommand{\elm}{electromagnetic~}
\renewcommand{\kappa}{\varkappa}
\newcommand{\om}{\omega}\newcommand{\Om}{\omega_{p}}\newcommand{\al}{\alpha}
\begin{document}
\title{Surface plasmon for graphene in the Dirac equation model}
\author{M. Bordag}
\affiliation{Universit\"{a}t Leipzig, Institute for Theoretical Physics, Germany}
\email{bordag@itp.uni-leipzig.de}
\date{\small  compiled \today}

\begin{abstract}
We consider single-layer plane graphene with electronic excitations described by the Dirac equation. Using a known representation of the polarization tensor in terms of the spinor loop we show the existence of surface modes, i.e., of undamped in time excitations of the \elm field, propagating along the graphene. These show up in the TE polarization and exist at zero temperature.
\end{abstract}
%
\maketitle
\section{Introduction}
Surface plasmons are excitations of the \elm field traveling along the interface between two media or along a thin sheet, decreasing exponentially fast to both sides of the interface. These excitations have been studied in detail, both theoretically and experimentally, see, for instance \cite{Raether1988,Sarid2010}.
In the present paper we show the existence of such excitations on graphene. The latter enjoys at present much attention because of its quite special properties, see, for example, \cite{aber01-59-2619}. The electronic excitations, primarily  responsible for  the interaction with the \elm field, are best described by a Dirac equation with correspondingly chosen parameters \cite{neto09-81-109}. This holds, at least, near the Dirac points. The response of this model to electric and magnetic fields was convincingly investigated in \cite{bene09-42-275401}. For a recent introduction to this model see \cite{fial12-27-1260007}.

Using a random-phase approximation with this model,  surface modes on graphene were investigated  in \cite{vafe06-97-266406}. It was found that the existence of such modes is due to thermal excitations and that these do not exist at $T=0$. In a number of papers this result was refined including, for example, electron-electron interaction \cite{herb08-100-046403}. In \cite{hans08-103-064302,mikh07-99-016803} surface plasmons were found in a conductivity approach.

In general, surface modes exist on the interface between two dielectric media with permittivities $\ep_1$ and $\ep_2$, in case the relations $\ep_1\ep_2<0$ and $\ep_1+\ep_2<0$ hold simultaneously \cite{LL}. For the surface of metal, described by the  plasma model with
\be\label{ep}\ep^{\rm pl}=1-\frac{\Om^2}{\om^2},
\ee
where $\Om$ is the plasma frequency, these relations are satisfied for $\om\le\Om/2$. Such modes are known also on an infinitely thin metal sheet described by plasma confined to a plane, which may serve as a hydrodynamic model for graphene \cite{BV}. In this case, the response to the \elm field is described by matching conditions the field has to satisfy across the sheet. These conditions are quite simple and result in the reflection coefficients
\be\label{rhy}r_{\rm TE}^{\rm hy}=\frac{-1}{1-\frac{i c q}{\Om}},\qquad
            r_{\rm TM}^{\rm hy}=\frac{1}{1+\frac{\om^2}{i c q\Om}}
\ee
(the superscript 'hy' denotes the hydrodynamic model, $c$ is the speed of light), for a standard scattering setup. Thereby the \elm field is separated into the usual polarizations, TE and TM, with plane wave amplitudes $\Phi(t,\mathbf{r})\sim \exp(-i\om t+i \mathbf{k}_{||}\mathbf{x}_{||})\Phi(z)$,
\be\label{Fiz}\Phi(z)=\left\{\begin{array}{lc}e^{i qz}+r e^{-iqz},&(z<0),\\
                                            t e^{i qz},&(z>0) .
                            \end{array}\right.
\ee
Here the $z$-axis is perpendicular to the sheet, $\mathbf{k}_{||}$ is the in-plane wave number and $q$ is the wave number   perpendicular to the sheet. The Maxwell equations give
\be\label{Meq}\om^2=c^2({k}_{||}^2+q^2)
\ee
with $k_{||}=|\mathbf{k}_{||}|$ and the matching conditions result in the reflection coefficients \Ref{rhy} (and in corresponding expressions for the transmission coefficient $t$). For comparison we note the corresponding formulas for the case of  dielectric media on both sides of the interface. In that case, the $z$-dependence of the amplitude is $\sim \exp(\pm i q_{1,2}z)$ with wave numbers $q_1$ and $q_2$ in the respective media having permittivities $\ep_1$ and $\ep_2$. The Maxwell equations give $\ep_{1,2}\om^2=c^2(\mathbf{k}_{||}^2+q_{1,2}^2)$ and the reflection coefficients are
\be\label{rpl}  r_{\rm TE}^{\rm pl}=\frac{q_1-q_2}{q_1+q_2},\qquad
                r_{\rm TM}^{\rm pl}=\frac{\ep_2 q_1-\ep_1 q_2}{\ep_2 q_1+\ep_1 q_2}.
\ee
A surface plasmon appears for  real frequency $\om$, real wave number $\mathbf{k}_{||}$ and imaginary wave number $q$,
\be\label{eta}q=i \eta
\ee
(or, correspondingly, imaginary $q_{1,2}$), with real $\eta$ which is the decay length of the amplitude in direction perpendicular to the surface, for a frequency giving the reflection coefficient a pole.

In such setup, the spectrum of the \elm field consists of scattering modes, having real all, $\om$, $\mathbf{k}_{||}$ and $q$, and the surface modes, in case these exist. For metals described by the plasma model and for graphene described by the hydrodynamic model, surface modes exist in the TM polarization.

It must be mentioned that the usage of the terminus 'surface plasmon' is not unique. Frequently these are also called surface polaritions and excitations with complex frequency $\om$, which are, strictly speaking, resonances,  are included too. We mention also that surface plasmons need for their existence at least one translational invariant direction, i.e., at least the surface of a cylinder. If the surface does not have such direction, the poles of the reflection coefficient appear at complex frequency and the mode decays. It must be mentioned that such modes can be quite long living and are of interest for applications.

From the theoretical point of view, surface plasmons, as being part of the spectrum, are interesting per se. Special interest comes from the role of the surface plasmons in the Lifshitz formula describing the Casimir and van der Waals interaction, see \cite{intr05-94-110404,bord06-39-6173} in the non retarded and \cite{bord12-85-025005} in the retarded cases for metals described by the plasma model and \cite{bord06-39-6173} for the hydrodynamic model. Also these play an important role in the temperature corrections to the Lifshitz formula \cite{bord13}. Needless to mention their role in applications as nanophotonics.

In the next section we show the existence of such plasmons on  graphene at zero temperature. \\Further in this paper we use units with $\hbar=c=1$.

\section{Surface plasmons in the Dirac model}
The description of graphene by a Dirac equation is a model for the electronic excitations accounting for the linearity of their spectrum near the Fermi points.
In the language of quantum field theory one has to calculate a spinor loop in (2+1) dimensions in a metric
\be\label{gmunu}g_{\mu\nu}={\rm diag}(1,-v,-v),
\ee
where $v=1/300$ is the Fermi speed (in units of the speed of light), and to couple the emerging polarization tensor to the \elm field in (3+1) dimensions. This was done in a number of papers. We use the representation in \cite{bord09-80-245406} at $T=0$ (it was extended in \cite{fial11-84-035446} to $T\ne0$), since here the expressions for the reflection coefficients in terms of the polarization tensor were derived. At $T=0$, these   reflection coefficients are
\be\label{rDi}   r_{\rm TE}^{\rm Di}=\frac{-1}{1+Q_{\rm TE}^{-1}},\qquad
                r_{\rm TM}^{\rm Di}=\frac{ 1}{1+Q_{\rm TM}^{-1}}.
\ee
with
\be\label{QDi} Q_{\rm TE}=\frac{\al}{2p_{||}}\Phi(\tilde{p}),\qquad
                Q_{\rm TM}=\frac{\al p_{||}}{2\tilde{p}^2_{||}}\Phi(\tilde{p}),
\ee
with the Euclidean momenta
\be\label{p1}   p_{||}=\sqrt{p_4^2+\bar{p}^2}, \qquad  \tilde{p}_{||}=\sqrt{p_4^2+v^2\bar{p}^2}
\ee
and the function
\be\label{Fi} \Phi(\tilde{p})=\frac{N}{2\tilde{p}}
        \left(2m\tilde{p}-(\tilde{p}^2+m^2){\rm arctanh}\frac{\tilde{p}}{2m}\right)
\ee
at the Minkowskian '3'-momentum
\be\label{p2} \tilde{p}=\sqrt{p_0^2-v^2\bar{p}^2},
\ee
which is the relativistic invariant in the metric \Ref{gmunu}. These momenta are related by a Wick rotation, $p_0=ip_4$, and $\bar{p}$ is the spatial momentum. In \Ref{QDi}, $\al=e^2/(4\pi)$ is the fine structure constant and $N=4$ is the number of fermion species. The mass $m$ is the gap parameter.

The above momenta are related to the photon momenta in \Ref{Fiz} and \Ref{Meq} by
\be\label{p3}   p_0=\om, \qquad k_{||}=\bar{p}.
\ee
Thus
\be\label{p4}   \tilde{p}=\sqrt{\om^2-v^2k_{||}^2}
\ee
will be real for $\om\ge v k_{||}$ and the relations
\bea\label{p5}   p_{||}&=&\sqrt{-\om^2+k_{||}^2}=\eta  \nn \\
                \tilde{p}_{||}&=&i\sqrt{\om^2-v^2k_{||}^2}=i\tilde{p}
\eea
hold with real $\eta$ and $\tilde{p}$ implying the restriction $\om\le k_{||}$.
Using these notations, we rewrite
\be\label{QDi2}  Q_{\rm TE}=\frac{\al}{2\eta}\Phi(\tilde{p}),\qquad
                Q_{\rm TM}=-\frac{\al \eta}{2\tilde{p}^2}\Phi(\tilde{p}).
\ee
At this point the signs become important. First, we note
\be\label{Fineg}    \Phi(\tilde{p})\le 0
\ee
for $0\le\tilde{p}\le 2m$, which can be seen explicitly from \Ref{Fi}. It should be mentioned that this condition, together with the reality of $\eta$, defines the only region where $Q_{\rm TE}$ and $Q_{\rm TM}$ are real. The upper bound is the threshold of pair production. We emphasize that $\Phi(\tilde{p})$ takes negative values. This is due to the extra minus sign a spinor loop has from the statistics, in distinction from a bosonic loop. This sign makes the reflection coefficient of the TE polarization having a pole, and as a consequence, a surface plasmon appears in the spectrum.

For comparison we mention the hydrodynamic model. Using the same notations as in \Ref{QDi2} we have
\be\label{Qhy}  Q_{\rm TE}=\frac{\Om}{\tilde{p}},\qquad
                Q_{\rm TM}=-\frac{\tilde{p}\,\Om}{\om^2},
\ee
and the pole is possible only in the TM polarization.

Thus, in the Dirac model, we may have a pole in the reflection coefficient of the TE polarization. Its location is solution of the equation
\be\label{feq1} 1+Q_{\rm TE}=0.
\ee
Inserting from \Ref{QDi2} and \Ref{Fi}, this equation can be rewritten in the form
\be\label{feq2} \sqrt{k_{||}^2-\om^2}=
    2\al m\left[\left(\frac{2m}{\tilde{p}}
    +\frac{\tilde{p}}{2m}\right){\rm arctanh}\frac{\tilde{p}}{2m}-1\right],
\ee
where we used the first line in eq. \Ref{p5} for $\eta$ and have from the second line
\be\label{p6}   \tilde{p}=\sqrt{\om^2-v^2k_{||}^2}.
\ee
This equation defines a relation between $\om$ and $k_{||}$. Its solution,
\be\label{omsf}    \om_{\rm sf}(k_{||}),
\ee
is the dispersion relation of the surface plasmon. For this solution, from the reality of $\eta$  and   $\tilde{p}$  and using  eq. \Ref{p5}, the restriction
\be\label{ineq} vk_{||}\le\om_{\rm sf}(k_{||})\le k_{||}
\ee
follows. Further, the solutions does not exceed the threshold of pair creation, i.e., they satisfy
\be\label{thre}  \tilde{p}\le 2m,
\ee
since otherwise  $\Phi(\tilde{p})$ would not be real and the equation \Ref{feq2} could not have a real solution.

\begin{figure}\unitlength 1cm
 \begin{picture}(8,5)
  \put(0,0){\includegraphics[width=8 cm]{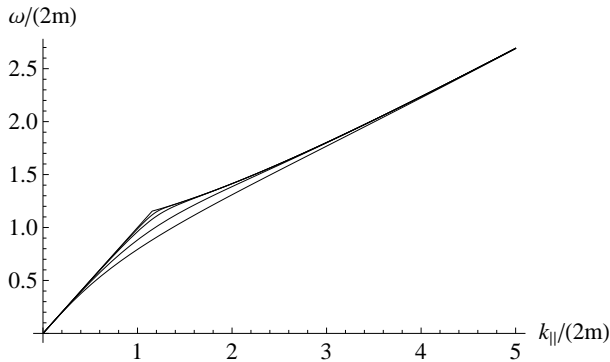}    }   \end{picture}
  \caption{The solutions of eq. \Ref{feq2} for $v=1/2$ and several values of the parameter ${\al}$. From bottom to top, the curves   correspond to ${\al}=1,\, 0.5,\,0.2,\,0.1,\,1/137$.}\label{fig1}
\end{figure}

\begin{figure}\unitlength 1cm
 \begin{picture}(8,5)
  \put(0,0){\includegraphics[width=8 cm]{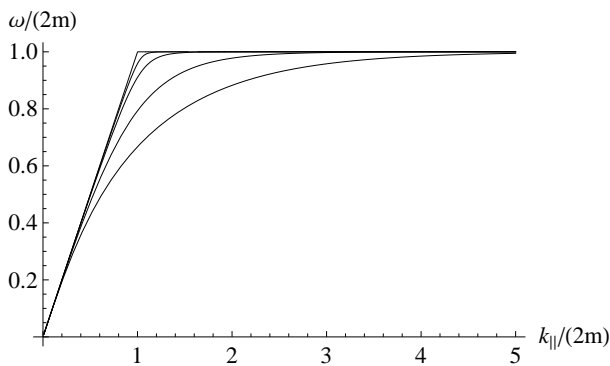}    }   \end{picture}
  \caption{The solutions of eq. \Ref{feq2} for $v=1/300$ and several values of the parameter ${\al}$. From bottom to top, the curves   correspond to ${\al}=1,\, 0.5,\,0.2,\,0.1,\,1/137$.}\label{fig2}
\end{figure}

The solution of eq. \Ref{feq2} exists for all $k_{||}$. It can be found numerically. Such solutions are shown in Fig. 1 for $v=1/2$ and in Fig 2 for $v=1/300$ for several values of $\al$.

The asymptotic behavior for small $k_{||}$ can be found by rewriting eq. \Ref{feq2} in the form
\be\label{eq1} \om=     \sqrt{k_{||}^2-\left(
    2\al m\left[\left(\frac{2m}{\tilde{p}}
    +\frac{\tilde{p}}{2m}\right){\rm arctanh}\frac{\tilde{p}}{2m}-1\right]\right)^2}
\ee
and iterating starting from inserting $\om=k_{||}$ in the right hand side. One obtains
\bea\label{sln1}\om_{\rm sf}(k_{||})&=&k_{||}-\frac{8}{9}\frac{\al^2(1-v^2)}{(2m)^3}\,k_{||}^3
\\\nn&& +\frac{32}{405}( 9-(35 +5 v^2)\al^2)\frac{\al^2(1-v^2)^3}{(2m)^4}\,k_{||}^5
    +\dots\,,
\eea
which is the expansion of the solution in powers of $k_{||}$. In order to get the behavior for large $k_{||}$, it is meaningful to rewrite the equation in the form
\begin{widetext}\be\label{eq2}  \om=
                \sqrt{v^2k_{||}^2+(2m)^2-\left[  \cosh
                \left(\frac{\sqrt{\om^2-v^2k_{||}^2}\left(\frac{1}{\al}\sqrt{k_{||}^2-\om^2}+1\right)}
                            {\om^2-v^2k_{||}^2+(2m)^2}     \right)   \right]^{-2}
                }.
\ee
\end{widetext}
It can be iterated by inserting $\om=\sqrt{v^2k_{||}^2+(2m)^2}$ in the right hand side. The solution is
\be\label{sln2} \om_{\rm sf}(k_{||})=\sqrt{v^2k_{||}^2+(2m)^2}+\dots\,,
\ee
where the dots denote contributions exponentially small for large $k_{||}$.

In this way, we have for both limiting cases a linear dispersion relation,
\be\label{lin}  \om_{\rm sf}(k_{||})=
        \left\{\begin{array}{rl}k_{||} & \mbox{~~~for} \to0,\\[8pt]
                                v k_{||} & \mbox{~~~for} \to\infty .   \end{array}\right.
\ee
The behavior in between these limiting cases, as can be seen from the figures, is smooth for any fixed values of the parameters $\al$ and $v$. For small $\al$, the curve has a quite sharp knee. For the physical values of the parameters it can be considered as consisting of two straight lines, crossing in $k_{||}=2m$.

The massless case corresponds to large both, $\om$ and $k_{||}$.     In that case we get from eq. \Ref{sln2}
\be\label{slnm0} \om_{\rm sf}(k_{||})_{|m=0}=v k_{||},
\ee
i.e., a linear dispersion relation. It must be mentioned that this limit cannot be performed in the function  $\Phi(\tilde{p})$, resulting from the polarization tensor for any finite $\tilde{p}$, because, as we know by hindsight, $\tilde{p}$ becomes small not to exceed the pair creation threshold \Ref{thre}.

\section{Conclusions}
We have shown the existence of a surface plasmon on graphene. We used the Dirac model, accounting for the spinor loop without further approximation. The surface plasmon shows up in the TE polarization which is due to the minus sign a Fermi loop has as compared to a bosonic one.

The surface plasmon appears at a frequency giving the reflection coefficient a pole. The corresponding equation, \Ref{feq2}, defines the frequency $\om_{\rm sf}(k_{||})$ as a unique function of the in-plane wave number $k_{||}$. This is the dispersion relation for this plasmon. For small and large $k_{||}$, eq. \Ref{lin}, and in the massless case, eq. \Ref{slnm0}, this relation is linear.

We found the surface plasmon at zero temperature, $T=0$. It is clear that it survives also  at finite temperature if that is sufficiently small. This is, because tempe\-ra\-ture corrections to $Q_{\rm TE}$ (as well as to $Q_{\rm TM}$), are always positive. These are small at sufficiently low temperature, but grow with increasing temperature until exceeding the negative $Q_{\rm TE}$.

We have to mention that the plasmons found here are different from those known in literature. The closest to ours are those in \cite{mikh07-99-016803} which exist in the TE polarization. However, the frequency condition used there (eq. (1) in \cite{mikh07-99-016803}) looks somehow oversimplified as compared with eq. \Ref{feq2}.

Finally we mention that the surface plasmon found here, has a good chance to exist on a carbon nano tube too. In the present paper we assumed a flat graphene sheet, neglecting the ripples. We did not estimate their influence. Also we did not discuss any interaction between the electrons in the graphene, or any interactions with the lattice which are beyond the considered model.

\section*{Acknowledgements}
The authors is indebted to Gabriel Barton for discussions a decade ago on surface plasmons on carbon structures.

\bibliographystyle{unsrt}
\bibliography{C:/Users/bordag/WORK/Literatur/bib/papers,C:/Users/bordag/WORK/Literatur/Bordag,C:/Users/bordag/WORK/Literatur/articoli}

\end{document}